\documentclass[a4p,11pt]{article}
\usepackage{epsfig}
\usepackage{cite}
\usepackage[margin=0.8in]{geometry}

\newcommand{\half}{\mbox{$\frac{1}{2}$}}

\newcommand{\dbyd}[2]{\ensuremath{\frac{{\rm d}{#1}}{ {\rm d}{#2}}}}

\newcommand{\dnbyddd}[5]{\ensuremath{\frac{{\rm d}^{#1}{#2}}{ {\rm d}{#3} {\rm
d}{#4} {\rm d}{#5} }}}

\newcommand{\prob}[1]{\ensuremath{ p\left({#1}\right)}}
\newcommand{\pcon}[2]{\ensuremath{ p\left({#1} \right|\left. {#2} \right)}}

\newcommand{\rpp}[1][]{\ensuremath{\rho^{#1}_{++}}}
\newcommand{\rmm}[1][]{\ensuremath{\rho^{#1}_{--}}}
\newcommand{\rzz}[1][]{\ensuremath{\rho^{#1}_{00}}}
\newcommand{\rpm}[1][]{\ensuremath{\rho^{#1}_{+-}}}
\newcommand{\rpz}[1][]{\ensuremath{\rho^{#1}_{+0}}}
\newcommand{\rmz}[1][]{\ensuremath{\rho^{#1}_{-0}}}

\newcommand{\KW}{KORALW}
\newcommand{\YFS}{YFSWW}
\newcommand{\KANDY}{KandY}

\newcommand{\CP}{\ensuremath{\mathrm{CP}}}
\newcommand{\CPTh}{\ensuremath{\mathrm{CP\hat{T}}}}

\newcommand{\T}{\ensuremath{\mathrm{T}}}
\newcommand{\Th}{\ensuremath{\mathrm{\hat{T}}}}
\newcommand{\sutwouone}{\ensuremath{SU(2)_{L}\times U(1)_{Y}}}

\newcommand{\W}{\ensuremath{\mathrm{W}}}
\newcommand{\Wm}{\ensuremath{\mathrm{W}^{-}}}
\newcommand{\Wp}{\ensuremath{\mathrm{W}^{+}}}

\newcommand{\WW}{\ensuremath{\mathrm{WW}}}

\newcommand{\Wln}{\ensuremath{\mathrm{W}\rightarrow \mathrm{\ell}\mathrm{\nu}}}
\newcommand{\Wqq}{\ensuremath{\mathrm{W}\rightarrow \mathrm{q}\mathrm{\bar{q}}^{\prime}}}

\newcommand{\Zzero}{$\mathrm{Z}^0$}

\newcommand{\lepton}{\ensuremath{\mathrm{\ell}}}

\newcommand{\epem}{\ensuremath{\mathrm{e}^{+}\mathrm{e}^{-}}}

\newcommand{\quark}{\ensuremath{\mathrm{q}}}

\newcommand{\GeV}{\mbox\rm{GeV}}
\newcommand{\mass}{\mbox{\rm{GeV/c}\ensuremath{^{2}}}}
\newcommand{\pb}{\mbox{\rm{pb}}}
\newcommand{\ipb}{\mbox{\rm{pb}}\ensuremath{^{-1}}}

\newcommand{\thw}{\ensuremath{\theta_{\W}}}
\newcommand{\cthw}{\ensuremath{\cos{\thw}}}
\newcommand{\thstar}[1]{\ensuremath{\theta^{*}_{#1}}}
\newcommand{\cthstar}[1]{\ensuremath{\cos{\thstar{#1}}}}
\newcommand{\cthstarsq}[1]{\ensuremath{\cos^{2}{\theta^{*}_{#1}}}}
\newcommand{\phistar}[1]{\ensuremath{\phi^{*}_{#1}}}

\newcommand{\qqen}{\ensuremath{\mathrm{q}\mathrm{\bar{q}^{\prime}}\mathrm{e}{\nu}}}
\newcommand{\qqmn}{\ensuremath{\mathrm{q}\mathrm{\bar{q}^{\prime}}\mathrm{\mu}{\nu}}} 
\newcommand{\qqtn}{\ensuremath{\mathrm{q}\mathrm{\bar{q}^{\prime}}\mathrm{\tau}{\nu}}}
\newcommand{\qqln}{\ensuremath{\mathrm{q}\mathrm{\bar{q}^{\prime}}\mathrm{\ell}{\nu}}}

\newcommand{\WWZ}{$\mathrm{WWZ}$}
\newcommand{\WWG}{$\mathrm{WW\gamma}$}

\newcommand{\eetoqqln}{\ensuremath{\mathrm{e}^+\mathrm{e}^- \rightarrow \mathrm{W}^+\mathrm{W}^- \rightarrow \mathrm{q}\mathrm{\bar{q}^{\prime}}\ell{\nu}_{\ell}}}

\newcommand{\twopho}{\ensuremath{\mathrm{e}^{+}\mathrm{e}^{-}\rightarrow\mathrm{e}^{+}\mathrm{e}^{-}\gamma\gamma\rightarrow\mathrm{e}^{+}\mathrm{e}^{-}\mathrm{q}\mathrm{\bar{q}^{\prime}}}}

\newcommand{\ZGqq}{\ensuremath{\mathrm{Z}^{0}/\gamma \rightarrow \mathrm{q}\mathrm{\bar{q}}}}

\newcommand{\etal}{\mbox{\it et al.}}

\newcommand{\com}{centre-of-mass}
\newcommand{\SM}{SM}
\newcommand{\TGC}{TGC}
\newcommand{\cc}{\ensuremath{\mbox{\rm CC}03}}

\newcommand{\PLB}[3]  {Phys.\ Lett.\ \textbf{B#1} (#2) #3}
\newcommand{\ZPC}[3]  {Z.\ Phys.\ \textbf{C#1} (#2) #3}
\newcommand{\EPC}[3]  {Eur.\ Phys.\ J.\ \textbf{C#1} (#2) #3}
\newcommand{\NIMA}[3] {Nucl.\ Instr.\ Meth.\ \textbf{A#1} (#2) #3}

\newcommand{\PRD}[3]  {Phys.\ Rev.\ \textbf{D#1} (#2) #3}
\newcommand{\NPB}[3]  {Nucl.\ Phys.\ \textbf{B#1} (#2) #3}

\newcommand{\IJMPA}[3]  {Int.\ J.\ Mod.\ Phys.\ \textbf{A#1} (#2) #3}
\newcommand{\CPC}[3]  {Comput.\ Phys.\ Commun.\ \textbf{#1} (#2) #3}

\begin{document}

\begin{titlepage}

\begin{center}
 {\large   EUROPEAN ORGANIZATION FOR NUCLEAR RESEARCH }
\end{center}

\bigskip

\begin{flushright}
       CERN-EP-2003-088  \\ 12th December 2003
\end{flushright}

\bigskip\bigskip\bigskip\bigskip\bigskip

\begin{center}
 {\huge\bf {\boldmath \W} Boson Polarisation at LEP2 }
\end{center}

\bigskip\bigskip

\begin{center}
{\LARGE The OPAL Collaboration }

\bigskip

\end{center}


\begin{center}
 {\large  Abstract}
\end{center}

Elements of the spin density matrix for \W\ bosons in \eetoqqln\ events
are measured from data recorded by the OPAL detector at LEP.
This information is used to calculate polarised differential cross-sections and to search for \CP-violating effects.
Results are presented for \W\ bosons produced in \epem\ collisions with
\com\ energies between 183~\GeV\ and 209~\GeV .
The average fraction of \W\ bosons that are longitudinally polarised is
found to be 
$(23.9\pm 2.1 \pm 1.1) \%$ compared to a Standard Model prediction of
$(23.9\pm0.1)\%$.
All results are consistent with \CP\ conservation.

\bigskip
\bigskip

\begin{center}
 {To be submitted to Physics Letters B.}
\end{center}

\end{titlepage}

\begin{center}{\Large        The OPAL Collaboration
}\end{center}\bigskip
\begin{center}{
G.\thinspace Abbiendi$^{  2}$,
C.\thinspace Ainsley$^{  5}$,
P.F.\thinspace {\AA}kesson$^{  3,  y}$,
G.\thinspace Alexander$^{ 22}$,
J.\thinspace Allison$^{ 16}$,
P.\thinspace Amaral$^{  9}$, 
G.\thinspace Anagnostou$^{  1}$,
K.J.\thinspace Anderson$^{  9}$,
S.\thinspace Arcelli$^{  2}$,
S.\thinspace Asai$^{ 23}$
% ...
}\end{center}\bigskip
\bigskip
$^{  1}$School of Physics and Astronomy, University of Birmingham,
Birmingham B15 2TT, UK
\newline
$^{  2}$Dipartimento di Fisica dell' Universit\`a di Bologna and INFN,
I-40126 Bologna, Italy
\newline
$^{  3}$Physikalisches Institut, Universit\"at Bonn,
D-53115 Bonn, Germany
\newline
% ...
$^{  5}$Cavendish Laboratory, Cambridge CB3 0HE, UK
\newline
% ...
$^{  9}$Enrico Fermi Institute and Department of Physics,
University of Chicago, Chicago IL 60637, USA
\newline
% ...
$^{ 16}$Department of Physics, Schuster Laboratory, The University,
Manchester M13 9PL, UK
\newline
% ...
$^{ 22}$Department of Physics and Astronomy, Tel Aviv University,
Tel Aviv 69978, Israel
\newline
$^{ 23}$International Centre for Elementary Particle Physics and
Department of Physics, University of Tokyo, Tokyo 113-0033, and
Kobe University, Kobe 657-8501, Japan
\newline
% ...
$^{  y}$ now at CERN
\newline
% ...

\newpage
\section{Introduction \label{section_intro}}
We present a spin density matrix (SDM) analysis~\cite{gounaris_sdm} of \W\ bosons pair-produced in \epem\ collisions at LEP using data collected by the OPAL collaboration with \com\ energies between 183~\GeV\ and 209~\GeV .

The Standard Model (\SM ) tree-level Feynman diagrams for \W\ pair production
at LEP are the $\mathit{s}$-channel diagrams with either a \Zzero\ or a photon
propagator and the $\mathit{t}$-channel neutrino exchange diagram.
Following the standard nomenclature, these three charged current diagrams are
collectively referred to as \cc~\cite{YR_cc03}.
The \WWZ\ and \WWG\ vertices in the $\mathit{s}$-channel Feynman diagrams represent triple gauge couplings (\TGC).

By measuring the spin state of the \W\ bosons we can investigate the physics of
the \TGC\ vertices and thereby test the \sutwouone\ gauge structure of the electroweak sector. 
The longitudinal helicity component of the spin is of particular interest as
it arises in the \SM\ through the electroweak symmetry-breaking mechanism which generates the mass of the \W .
In addition, comparisons of the spin states of the \Wm\ and \Wp\ are sensitive
to \CP-violating effects.
Such effects are absent both at tree-level and at the one-loop level in the \SM.

In order to probe the spin state of a \W\ boson, it is necessary to
reconstruct the directions in which its decay products are emitted.
The polar angle of the \Wm\ with respect to the electron beam direction is
denoted by \thw\ throughout this paper.
The polar and azimuthal angles of the outgoing fermion in the rest frame of
the parent \Wm\ are denoted by \thstar{f}\ and \phistar{f}\ respectively~\footnote{The axes of the right-handed coordinate system in the
parent \W\ rest frame are defined using the helicity axes convention such that
the $z^{*}$-axis is along the boost direction, $\hat{b}$,
from the laboratory frame, and the $y^{*}$-axis is in the direction
$\hat{\mathrm{e}}^{-*} \times \hat{b}$, where $\hat{\mathrm{e}}^{-*}$ is the
direction of the incoming electron beam.}.
The polar and azimuthal angles of the outgoing anti-fermion in the rest frame
of the parent \Wp\ are denoted by \thstar{\bar{f}}\ and \phistar{\bar{f}}\ respectively.

The SDM formalism is described in section~\ref{section_sdm} whilst the
practical implementation is explained in section~\ref{section_measure}.
Section~\ref{section_data} details the data samples and Monte Carlo
simulations used in this analysis. 
The sources of systematic uncertainty in the results are explained in section~\ref{sec_systematics}.
The measured fractions of longitudinally polarised \W\ bosons and other
related results are presented in section~\ref{section_results} and their
implications discussed in section~\ref{section_conclusion}. 
 
\section{Representation of the {\boldmath \W} boson spin state \label{section_sdm}}

\subsection{The spin density matrix \label{subsection_sdm}}

The angular distributions of the decay products of an ensemble of \W\ bosons
can be analysed to determine the composition of the average spin state, which
depends on the \com\ energy of the reaction, $\sqrt{\mathit{s}}$, and the
polar production angle of the \Wm, \thw.
It is convenient to represent this composition by the spin density
matrix, $\rho_{\tau\tau^{\prime}}$, defined as~\cite{gounaris_sdm}:
\begin{eqnarray}
 \rho^{\Wm}_{\tau\tau^{\prime}}(s,\cthw)&=& \frac{ \sum_{\lambda,
\lambda^{\prime}}
F^{(\lambda,\lambda^{\prime})}_{\tau}
(F^{(\lambda,\lambda^{\prime})}_{\tau^{\prime}})^{*} }{\sum_{\lambda,
\lambda^{\prime},
\tau} \left| F^{(\lambda,\lambda^{\prime})}_{\tau} \right|^{2} } \ \ , 
\end{eqnarray}
 where $F^{(\lambda,\lambda^{\prime})}_{\tau}$ is the helicity amplitude for producing a \Wm\
with helicity $\tau$ from an electron with helicity $\lambda$ and positron
with helicity $\lambda^{\prime}$.
As the \W\ boson has unit spin, $\tau$ can take the values $+1$, $0$ or $-1$.
The SDM for the \Wp\ is defined analogously in terms of the \Wp\ helicity
amplitudes.

The SDM is an Hermitian matrix with unit trace, and is fully described by eight free parameters.
The elements lying on its major diagonal are the probabilities of observing a
\W\ boson in each of the three possible helicity states and are therefore
positive as well as purely real.
The real and imaginary parts of the off-diagonal terms measure the interference between the helicity amplitudes.

\subsection{Projection operators \label{subsection_projop}}
Assuming a $\mathrm{V}-\mathrm{A}$ structure for the \Wm\ coupling to fermions, the
expected angular distribution for massless fermions in the rest frame of the
parent \Wm\ is given in terms of the diagonal elements of the
SDM~\cite{YR_tgc} by:
\begin{eqnarray}
\frac{1}{\sigma}\dbyd{\sigma}{ \cthstar{f}}&=&
\rho_{--}\frac{3}{8}\left(1+\cthstar{f} \right)^{2} +
\rho_{00}\frac{3}{4}\sin^{2}\theta^{*}_{f} \nonumber \\
 & & + \rho_{++}\frac{3}{8}\left(1-\cthstar{f} \right)^{2} \ \ . \label{eqn_helfunc}
\end{eqnarray}

In this paper, $\sigma$ is the cross-section for the process \eetoqqln.
In practice, the ratios of the masses of the \SM\ leptons and quarks to the mass of the \W\ boson are sufficiently small that deviations from equation
(\ref{eqn_helfunc}) are negligible compared to the statistical precision of the
measurements being made\footnote{The top quark is too massive to be produced from on-shell \W\ bosons and the production of bottom quarks is highly
suppressed.}. 
Hence, it is possible to extract the diagonal elements of the SDM by fitting this function to the \cthstar{f}\ distribution obtained from the data.  
Such an analysis has been published by the L3 collaboration~\cite{L3}.

As in previous OPAL analyses~\cite{opal_pr323,opal_pr260}, we use the alternative method of
constructing projection operators, $\Lambda_{\tau\tau^{\prime}}$, which
satisfy: 
\begin{eqnarray}
\rho^{\Wm}_{\tau\tau^{\prime}}(s,\cthw)&=& \frac{\int
\dnbyddd{3}{\sigma}{ \cthw}{ \cthstar{f}}{ \phistar{f}} \cdot
\Lambda_{\tau\tau^{\prime}} {\rm d}\!\cthstar{f} {\rm d} \phistar{f} }{
\dbyd{\sigma}{ \cthw} } \ \ .  
\end{eqnarray}

Equations (\ref{eqn_projmm}) to (\ref{eqn_projpp}), listed below, give the projection operators used to extract the diagonal elements of the \Wm\ SDM.
They depend only on the polar production angle of the fermion,
$\theta_{f}^{*}$.
The projection operators corresponding to the off-diagonal elements of the
\Wm\ SDM are given in equations (\ref{eqn_projpm}) to (\ref{eqn_projmz}) and
have azimuthal angular dependence.

\begin{eqnarray}
\Lambda_{--}&=& \half (5\cthstarsq{f}+2\cthstar{f}-1) \label{eqn_projmm} \\
\Lambda_{00}&=& 2-5\cthstarsq{f} \label{eqn_projzz}  \\
\Lambda_{++}&=& \half (5\cthstarsq{f}-2\cthstar{f}-1) \label{eqn_projpp} 
\end{eqnarray}
\begin{eqnarray}
\Lambda_{+-}&=& 2e^{2i\phistar{f}} \label{eqn_projpm} \\
\Lambda_{+0}&=& \frac{-8}{3\pi\sqrt{2}}\cdot\left(1-4\cthstar{f}\right)e^{-i\phistar{f}}\label{eqn_projpz}  \\
\Lambda_{-0}&=&\frac{-8}{3\pi\sqrt{2}}\cdot\left(1+4\cthstar{f}\right)e^{i\phistar{f}} \label{eqn_projmz} 
\end{eqnarray}

The operators for the \Wp\ can be obtained by replacing \cthstar{f}\ and
\phistar{f}\ by \cthstar{\bar{f}}\ and \phistar{\bar{f}}. 

\subsection{Polarised cross-sections \label{subsection_polxs}}
The longitudinally and transversely polarised differential cross-sections for
\W\ production are given in terms of the SDM elements by~\cite{gounaris_sdm}:

\begin{eqnarray}
\dbyd{\sigma_{L}}{ \cthw}&=&  \rho_{00} \cdot \dbyd{\sigma}{ \cthw} \label{eqn_diffxslong} \\  
\dbyd{\sigma_{T}}{ \cthw}&=& \left(\rho_{++} + \rho_{--}\right) \cdot
\dbyd{\sigma}{ \cthw} \ \ .  \label{eqn_diffxstrans}
\end{eqnarray}

The electric charge of the charged lepton in a \qqln\ event can be reliably
reconstructed so that there is no ambiguity in
determining \cthstar{f}\ or \phistar{f}\ for use in the projection operators.
The charge of a quark which originates a hadronic jet is not readily
accessible, so the measured angular distributions for hadronically decaying
\W\ bosons are folded such that \cthstar{f}\ lies between 0 and 1 and
\phistar{f}\ lies between 0 and $\pi$.
Although neither $\rho_{++}$ nor $\rho_{--}$ can be measured individually
after this folding, their sum is unchanged and the polarised differential cross-sections can still be evaluated. 

The total polarised cross-sections are obtained by integrating the
differential cross-sections with respect to \cthw.

\subsection{Effects of {\boldmath \CP} violation \label{subsection_cptest}}
Much of the sensitivity of \W\ pair production to \CP-violating interactions
is contained in the distributions of the azimuthal angles \phistar{f} and \phistar{\bar{f}}.
Both azimuthal angular distributions are symmetric about zero at tree-level in
the \SM.
The presence of a \CP-violating phase at the \TGC\ vertex would, in general,
shift the distributions to introduce an asymmetry.
This effect can be measured from the off-diagonal elements of the SDM, as described below.

Under the assumption of \CP\ invariance, the SDM for the \Wm\ and the SDM for
the \Wp\ are related by~\cite{gounaris_cp}:
\begin{eqnarray}
\rho^{\Wm}_{\tau\tau^{\prime}}&=&\rho^{\Wp}_{-\tau -\tau^{\prime}} \ \ . 
\label{eqn_cp}
\end{eqnarray} 

The time-reversal operator \T\ can be approximated by the pseudo time-reversal
operator \Th\ which transforms the helicity amplitudes into their complex
conjugates rather than interchanging their initial and final states.
At tree-level, the effect of the pseudo time-reversal operator is exactly equivalent to the effect of the true time reversal operator~\cite{chang_cp, hagiwara_87}.
Under the assumption of \CPTh\ invariance, the SDM for the \Wm\ and the SDM for the \Wp\ are related by:

\begin{eqnarray}
\rho^{\Wm}_{\tau\tau^{\prime}}&=&\left(\rho^{\Wp}_{-\tau
-\tau^{\prime}}\right)^{*} \label{eqn_cpt} \ \ . 
\end{eqnarray} 

It follows that tree-level \CP\ non-conserving effects will only violate the
imaginary part of equation (\ref{eqn_cp}). 
This motivates the construction of $\sigma^{\Wm}_{\tau\tau^{\prime}}$ and
$\sigma^{\Wp}_{\tau\tau^{\prime}}$ which are defined below in terms of the imaginary parts of the off-diagonal
SDM elements and are measured in units of cross-section:
 
\begin{eqnarray}
\sigma^{\Wm}_{\tau\tau^{\prime}}&=&\int_{-1}^{+1}Im\left\{\rho^{\Wm}_{\tau\tau^{\prime}}\right\}
\cdot \dbyd{\sigma}{ \cthw} {\rm d} \cthw \label{eqn_wminuscpxsec}\\
\sigma^{\Wp}_{\tau\tau^{\prime}}&=&\int_{-1}^{+1}
Im\left\{\rho^{\Wp}_{\tau\tau^{\prime}}\right\} \cdot \dbyd{\sigma}{ \cthw}
{\rm d} \cthw \label{eqn_wpluscpxsec} \ \ . 
\end{eqnarray}

From these quantities we form experimentally accessible \CP-odd observables:
\begin{eqnarray}
\Delta^{\CP}_{+-}&=&\sigma^{\Wm}_{+-}-\sigma^{\Wp}_{-+}   \\
\Delta^{\CP}_{+0}&=&\sigma^{\Wm}_{+0}-\sigma^{\Wp}_{-0}   \\
\Delta^{\CP}_{-0}&=&\sigma^{\Wm}_{-0}-\sigma^{\Wp}_{+0}   \ \ . 
\end{eqnarray}

We also construct \CPTh-odd observables sensitive to the presence of
loop effects: 
\begin{eqnarray}
\Delta^{\CPTh}_{+-}&=&\sigma^{\Wm}_{+-}+\sigma^{\Wp}_{-+}  \\
\Delta^{\CPTh}_{+0}&=&\sigma^{\Wm}_{+0}+\sigma^{\Wp}_{-0}  \\
\Delta^{\CPTh}_{-0}&=&\sigma^{\Wm}_{-0}+\sigma^{\Wp}_{+0} \ \ . 
\end{eqnarray}

No assumptions about the form of the \TGC\ vertices are necessary to extract these
observables from the data.
Hence this study is complementary to the \CP-violating \TGC\ parameter
measurements previously published by ALEPH and OPAL~\cite{opal_pr323,
aleph_2000_015}.

\section{Data and Monte Carlo simulations \label{section_data}}

\subsection{Data sample}
This analysis used data collected by the OPAL detector~\cite{OPAL} at LEP during the years 1997 to 2000.
The data were collected with \com\ energies clustered around eight nominal
energy points: 183~\GeV, 189~\GeV, 192~\GeV, 196~\GeV, 200~\GeV, 202~\GeV,
205~\GeV\ and 207~\GeV. 

Events with \qqln\ final states originating from a pair of \W\ bosons, where
the charged lepton can be either an electron, muon or $\tau$-lepton, are referred
to as {\it signal} in the remainder of the paper.
The final state can additionally contain any number of photons.
Events of all other types are referred to as {\it background}.  
Only those data events compatible with the signal definition were used to form
the SDM.

The luminosity-weighted mean \com\ energies and integrated
luminosities of the data samples, as obtained from measurements of small angle
Bhabha events in the silicon tungsten forward calorimeter~\cite{opal_sw}, are listed in table~\ref{table_lumi}.
The total integrated luminosity was 678.5~\ipb, which corresponds to a \SM\ 
prediction of approximately 5000 signal events being produced.
The eight data samples were analysed separately and the results from each of
these analyses were then combined using the method described in section~\ref{section_results}.

\begin{table}[h]
\begin{center}
\begin{tabular}{|c||c|c|c|} \hline 
$\sqrt{\mathit{s}}$ (\GeV)& $\int{\mathcal{L}dt}$(\ipb)&observed events&expected events \\ \hline
182.7&57.40&329&331.0\\ 
188.6&183.0&1090&1124.7\\ 
191.6&29.3&170&182.7\\ 
195.5&76.4&511&483.2\\ 
199.5&76.6&457&479.4\\ 
201.6&37.7&242&237.4\\ 
204.9&81.6&482&516.0\\ 
206.6&136.5&895&862.7\\ \hline 
Total&678.5&4176&4214.5\\ \hline
\end{tabular}
\caption{Mean \com\ energies and integrated luminosity values for the
data. The number of data events passing the
event selection detailed in section~\ref{section_selection} and the
expected number of events as calculated using the Monte Carlo simulations
listed in section~\ref{section_montecarlo} are also shown.\label{table_lumi}}
\end{center}
\end{table}
 
\subsection{Monte Carlo simulations\label{section_montecarlo}}

Samples of simulated data events with four-fermion final states consistent with
having been produced via a pair of \W\ bosons were generated using the
\KANDY\ Monte Carlo (MC) generator formed from the \YFS 3~\cite{yfs} and \KW~1.51~\cite{koralw} software packages.
These events were weighted by factors calculated using \KANDY\
to provide \cc\ signal samples (the \cc\ Feynman
diagrams are described in section~\ref{section_intro}).
Four-fermion final states inconsistent with having been produced 
via a pair of \W\ bosons were generated using \KW~1.42~\cite{koralw1.42}.
Additional samples of four-fermion and \cc\ events used in studies of
systematic effects were also generated using \KW~1.42.
The EXCALIBUR~\cite{excalibur} MC
generator was used to re-weight four-fermion events for parts of the systematic error analysis.
The \KANDY\ and \KW\ samples used in the main analysis were hadronised using JETSET~\cite{jetset}.
To estimate the fragmentation and hadronisation systematic uncertainties, 
HERWIG~6.2~\cite{herwig} and ARIADNE~4.11~\cite{ariadne} were used as
alternatives to JETSET for simulating hadronisation in some \KW\ samples.

In addition to four-fermion events, only 
quark-pair or two-photon events were found to be significant sources of
background for the event selection described in
section~\ref{section_selection}.
Samples of \ZGqq\ events were generated by KK2F~\cite{kk2f} and hadronised by
JETSET whilst multi-peripheral two-photon processes with hadronic final states (\twopho) were simulated by HERWIG.

The MC samples were processed by the full OPAL simulation program~\cite{gopal}
and then reconstructed in the same way as the data.

\section{Measurement of the \boldmath{\W} boson spin state \label{section_measure}}

\subsection{Event selection and reconstruction\label{section_selection}}

The selection algorithm applied to the data to identify \qqln\ candidate
events had four parts: the pre-selection, the likelihood selection, the
kinematic fit used to derive estimates of the momentum vectors of the four
fermions from the \W\ decays, and the final selection.
Details of the pre-selection and likelihood selection have been published previously in~\cite{opal_pr321,opal_pr260} and were based on the 172~\GeV\ \qqln\ selection described in appendix A of~\cite{opal_pr217}.
Details of the kinematic fits and final selection are given in~\cite{opal_pr324}.
The number of events passing the full selection is shown in
table~\ref{table_lumi}.

For each selected \qqln\ candidate event, the fitted momentum vectors of the four fermions were used to calculate
the production and decay angles of the \W\ bosons required for the remainder of
the analysis.
The events were then divided into eight \cthw\ bins of equal width.
The true number of signal events produced in each bin was estimated by:
\begin{eqnarray}
N_{k}&=&\sum^{n_{k}}_{i=1}{\left(\frac{p}{\epsilon}\right)^{(i)}} \ \ , \label{eq_statdiffxs}
\end{eqnarray}
where the sum is over all $n_{k}$ data events reconstructed in the bin and 
$p/ \epsilon$ is a detector correction factor (defined in section~\ref{sec_detcorr}) which allows for variations with the polar and azimuthal angles of the efficiency, purity and angular resolutions.
The unpolarised differential cross-section needed to evaluate
equations~(\ref{eqn_diffxslong}), (\ref{eqn_diffxstrans}),
(\ref{eqn_wminuscpxsec}) and (\ref{eqn_wpluscpxsec})  was estimated by dividing these numbers by the luminosities given in table~\ref{table_lumi}. 

Using the same notation as above, the statistical estimators for the SDM elements were given by:
\begin{eqnarray}
\rho^{k}_{\tau\tau^{\prime}}&=&\frac{1}{N_{k}}\sum^{n_{k}}_{i=1}{\left(\frac{p}{\epsilon}\right)^{(i)}.\
\Lambda^{(i)}_{\tau\tau^{\prime}} } \label{eq_statrho} \ \ ,
\end{eqnarray}
The estimators for the diagonal elements of the SDM were not explicitly
constrained to lie between 0 and 1, and hence could have unphysical values
due to statistical fluctuations in the data.
In addition, the detector correction treated $p/ \epsilon$ as statistically
independent in each angular bin.
Therefore, the values of the estimators in different \cthw\ bins were statistically uncorrelated.

As the \CP\ and \CPTh\ symmetry tests rely on measuring asymmetries in the
azimuthal angular distributions, the folded distributions obtained from
hadronically decaying \W\ bosons were not used in evaluating the off-diagonal
elements of the SDM.

\subsection{Detector correction \label{sec_detcorr}}
The MC samples listed in section~\ref{section_montecarlo} were used to estimate the
efficiency and purity of the event selection and the angular resolution with which the directions of the \W\ bosons and their decay products were reconstructed in the detector. 
The overall selection efficiency varied between $78\%$ and $81\%$, and the purity varied between $93\%$ and $95\%$ depending on the \com\ energy.
In addition, the efficiency varied between $44\%$ and $92\%$, and the purity
varied between $15\%$ and $100\%$ in the \cthw-\cthstar{l}\ plane.
Variations of the efficiency and purity with \cthstar{q}\ and
\phistar{l}\ were smaller but significant.
The angular resolution was dominated by the measurement uncertainties in the
four-momenta of the hadronically decaying \W\ bosons, but additionally included
a small contribution from the $3\%$ of charged leptons which were assigned an incorrect charge (mostly in the \qqtn\ channel).
As the generator-level MC values of \thw\ were obtained after boosting to the
\com\ frame of the four fermions whereas the reconstructed values 
were measured in the lab frame, the angular resolution was 
sensitive to initial-state radiation (ISR) effects.

After reconstruction, each data event was scaled by a detector correction
factor ($p/ \epsilon$ in equation (\ref{eq_statrho})) given by the purity divided by an efficiency-like scaling factor.
The scaling factor was defined as the ratio of signal MC events reconstructed
in a given angular bin divided by the number of signal MC events generated in that bin, and hence included the effects of both the efficiency and
angular resolution under the assumption that the shapes of the MC angular
distributions closely approximated those of the data.
Approximately $16\%$ of the MC events which passed the selection were
reconstructed outside of the \cthw\ bin in which they were generated, where
the exact fraction depended on the \com\ energy.  
In studies using the full unfolding procedures described
in~\cite{vato,agostini}, the bin-to-bin migration led to correlations as high as $40\%$ between the statistical errors in neighbouring \cthw\ bins.
As the SDM estimators of section~\ref{section_selection} are statistically
uncorrelated, special care should be exercised if using the results of this
analysis in fitting procedures.

The numbers of angular bins used to parameterise the detector
correction for leptonically and hadronically decaying \W\ bosons and for diagonal
and off-diagonal elements of the SDM are summarised in table~\ref{table_binning}. 
These were chosen to make best use of the MC statistics whilst reducing
possible bias effects. 
The corrections for positively and negatively charged \W\ bosons were
combined by making use of the approximate \CP\ invariance of both the \SM\ MC and the response of the OPAL detector~\cite{opal_pr191}. 

\begin{table}[h]
\begin{center}
\begin{tabular}{|c|c|c|c|c|c|} \hline 
SDM elements\rule[-2mm]{0mm}{5mm}&Decay mode&\cthw&\cthstar{\lepton}&\phistar{\lepton}&\cthstar{\quark}\\ \hline
\rmm, \rpp, \rzz & \Wln& 8 & 20 & - & - \\ \cline{2-6}
                 & \Wqq& 8 & -  & - & 10 \\ \hline
\rpm, \rpz, \rmz& \Wln& 8 & 5  & 5 & - \\ \cline{2-6}
                 & \Wqq& - & - &  - & - \\ \hline
\end{tabular}
\caption{The numbers of bins, of equal width, used to parameterise the detector
correction. The hadronically decaying \W\ bosons were
not used to measure the off-diagonal SDM elements. \label{table_binning}}
\end{center}
\end{table}

\subsection{Bias correction \label{sec_biascorr}}
The method employed to compensate for detector effects, described in section~\ref{sec_detcorr}, tended to bias the final result towards the Standard Model prediction. 
In order to correct for this effect in the \W\ polarisation measurement, a
bias correction was applied to each diagonal element of the SDM and to the
\cthw\ distribution.
For each \cthw\ bin, the biases in the measured values of \rzz, \rmm\ and in
the number of reconstructed events, $N$, were calculated from re-weighted MC samples using a grid of 40 equally spaced values of \rzz\ and 40 equally spaced values of \rmm\ distributed between 0 and 1. 
The bias correction, $\bar{b}$, was taken to be the average bias as shown in
equation (\ref{eq_bias}), where the sum is over the points on the grid.
The vector $\vec{\rho_{i}}$ represents the values of \rzz\ and \rmm\ for the $i$'th re-weighted MC sample.        
The vector $\vec{r}$ represents the values measured from the data sample with error matrix $R$.
The bias in the variable of interest (\rzz, \rmm\ or $N$) is denoted by $b_{i}$.
The bias correction for \rpp\ was calculated using the bias corrections for
\rzz\ and \rmm\ and the normalisation constraint $\rpp+\rmm+\rzz=1$.
 
\begin{eqnarray}
\bar{b}&=&\sum_{i}{ \pcon{\vec{\rho_{i}}}{\vec{r}}\: .\: b_{i} }
\label{eq_bias} \nonumber \\
       &=&\sum_{i}{ \frac{\pcon{\vec{r}}{\vec{\rho_{i}}} .
\prob{\vec{\rho_{i}}}}{\prob{\vec{r}}}\: .\: b_{i} } \nonumber \\
       &\propto&\sum_{i}{ \exp\left[ {\frac{1}{2}\left( \vec{r} - \vec{\rho_{i}}
\right)R^{-1}\left( \vec{r} - \vec{\rho_{i}} \right)}\right] \: .\: b_{i} }          \label{eq_bias3} 
\end{eqnarray}

Equation (\ref{eq_bias3}) was derived from Bayes' theorem assuming 
a uniform prior probability distribution for $\vec{\rho_{i}}$.
The values of $\bar{b}$ for elements of the SDM typically had a magnitude less than 0.1.
The uncertainty on each bias correction was estimated by the standard deviation of the biases, $b_{i}$.
These error estimates were typically smaller than 0.2, and are included in the statistical errors of the results in section~\ref{section_results}.

As the bias varied significantly over the range of measured values
spanned by the statistical errors on the data, the uncertainty on the
correction was often larger than the correction itself.
The shifts in the fraction of longitudinally polarised \W\ bosons due to the
introduction of the bias corrections are shown in table~\ref{table_bias} where the uncertainty in each shift has been evaluated by varying the bias corrections within their errors.
In each case, the shift due to the bias corrections is smaller than the
statistical error on the measurement (see table~\ref{table_wpol}).
Tests with MC samples were used to show that this bias correction procedure
results in unbiased estimators which give Gaussian pull distributions.

\begin{table}[h]
\begin{center}
\begin{tabular}{|c||r|r|} \cline{2-3}
\multicolumn{1}{c|}{}&\multicolumn{2}{c|}{Shifts (\%) }\\ \hline
\multicolumn{1}{|c||}{$\sqrt{\mathit{s}}$ (\GeV)}&\Wln&\Wqq\\ \hline
183&$-0.8 \pm 1.7$ &$ 3.0 \pm 1.9$\\ \hline
189&$-1.1 \pm 1.1$ &$-0.8 \pm 1.2$ \\ \hline
192&$ 3.7 \pm 2.4$ &$ 2.0 \pm 1.9$\\ \hline
196&$-0.6 \pm 1.3$ &$-0.5 \pm 1.4$\\ \hline
200&$ 1.9 \pm 1.0$ &$ 2.7 \pm 2.2$\\ \hline
202&$-0.1 \pm 1.6$ &$ 2.5 \pm 2.4$\\ \hline
205&$ 0.6 \pm 1.5$ &$ 2.6 \pm 2.2$\\ \hline
207&$-1.0 \pm 1.1$ &$-0.3 \pm 1.6$\\ \hline
\end{tabular}
\caption{Shifts in the percentage of longitudinally polarised \W\ bosons due
to the bias corrections applied at each \com\ energy.  \label{table_bias}}
\end{center}
\end{table}
 
The computer processing time required to extend the bias correction to the off-diagonal elements of the SDM was prohibitive and so the simpler method described in section~\ref{sec_systematics} was used as an adequate alternative.

\section{Systematic errors \label{sec_systematics}}
There are uncertainties in the shape of the MC angular distributions due to 
the measurement errors on the parameters in the \SM\ (such as the mass of the
\W\ boson), to the incomplete description of non-perturbative physics effects in MC generators and to the simplifying assumptions made in modelling the detector response.
This leads to uncertainties in the detector corrections applied to the data and gives systematic errors on the final results.
For each source of systematic uncertainty, the full analysis of the data was
repeated using a range of different MC samples (or a single MC sample re-weighted
appropriately) to form the detector correction.
The full difference between the results obtained using each of
the detector corrections in turn was assigned as the error.
The total systematic error on each measured quantity was calculated by summing
the errors associated with each source of uncertainty in quadrature.
The error sources are listed below and their contributions to the total
systematic error for the luminosity-weighted average fraction of
longitudinally polarised \W\ bosons are shown in table~\ref{table_systematic}. 

\begin{enumerate}

\item The effects of the uncertainties in the modelling of each of the
different background MC samples were evaluated by varying the contribution from 
the two-photon samples by a factor of two and the contribution from the
four-fermion background samples by a factor of 1.2.
These factors were obtained using the method described in~\cite{opal_pr321}. 
Samples of \ZGqq\ events hadronised by JETSET were also compared
to samples hadronised by HERWIG. 

\item Unlike the MC generators used in the previous SDM analyses published by
the OPAL collaboration, the \KANDY\ MC generator includes a full treatment
of $\mathcal{O}(\alpha)$ electroweak loop corrections for the \cc\ diagrams
(using a double-pole approximation). 
The effect of the missing higher order corrections was estimated by
re-weighting the MC samples to remove $\mathcal{O}(\alpha)$ next-to-leading
electroweak corrections and the screened Coulomb correction.

\item The error due to uncertainty in the modelling of the jet fragmentation
was estimated by comparing the results obtained using \WW\ MC samples
hadronised via JETSET, HERWIG and ARIADNE.
MC samples were generated at 189~\GeV, 200~\GeV\ and 206~\GeV, and used to
interpolate the systematic errors at all eight nominal \com\ energies.
 
\item The \W\ mass obtained from Tevatron and UA2 data is 
$80.454 \pm 0.060~\mass$~\cite{PDG}.
The systematic error due to the difference between this mass and the \W\
mass used in generating the \KANDY\ MC samples ($80.33~\mass$) was estimated by
re-weighting the \KANDY\ samples using the EXCALIBUR MC generator.
The LEP measurements of the \W\ mass were not used to evaluate the error, as
they implicitly assume that the \W\ pairs are produced via the Standard Model mechanism.

\item The \KANDY\ MC generator includes an $\mathcal{O}(\alpha^{3})$ treatment of initial state radiation (ISR). The effect of missing higher order diagrams was
estimated by re-weighting the MC samples to use an $\mathcal{O}(\alpha)$ ISR
treatment. 

\item The modelling of the OPAL detector's response to hadrons and leptons was compared to that seen in the data using events collected at the \Zzero\ peak.
The angles and energies of the jets and charged leptons in the reconstructed MC were smeared to give better agreement with the \Zzero\ data.
The error associated with the uncertainty on the smearing was evaluated by
varying the smearing within its statistical error.
A fuller account of this procedure can be found in~\cite{opal_pr320}.
 
\item Some areas of the \W\ boson production and decay phase space were
sparsely populated by MC and data events.
The effect of the limited MC statistics was evaluated by smearing the efficiency and purity corrections by their statistical errors. 

\item Biases in the off-diagonal elements of the SDM due to the form of the
detector correction were not explicitly removed. Instead the \KANDY\
MC samples used to calculate the detector correction were re-weighted to
simulate anomalous values of the \CP-violating \TGC\ parameters which alter the shape of the \phistar{f}\ and \phistar{\bar{f}}\
angular distributions. The \TGC\ parameters were varied by one standard
deviation of their measured values from the OPAL 189~\GeV\ analysis~\cite{opal_pr323}.
The bias associated with the finite width of the angular bins into which the
detector correction was divided was also taken into account.

\end{enumerate}

\begin{table}[h]
\begin{center}
\begin{tabular}{|l|c|c|} \cline{2-3}
\multicolumn{1}{c|}{}&\Wln&\Wqq\\ \hline
Two-photon MC$^{\dagger}$  &0.10&0.08 \\ \hline
\ZGqq\ MC$^{\dagger}$      &0.30&0.32 \\ \hline
Four-fermion MC$^{\dagger}$&0.19&0.03 \\ \hline
$\mathcal{O}(\alpha)$ radiation$^{\dagger}$&0.23&0.18\\ \hline
Hadronisation$^{\dagger}$ &0.38&1.23 \\ \hline
$m_{W}$$^{\dagger}$       &0.13&1.07 \\ \hline
ISR$^{\dagger}$           &0.01&0.01 \\ \hline
Detector Response         &0.19&0.06 \\ \hline
MC statistics             &0.29&0.18 \\ \hline \hline
Total                     &0.68&1.68 \\ \hline 
\end{tabular}
\caption{The systematic errors for the luminosity-weighted average percentage
of longitudinally polarised \W\ bosons.
The error sources appear in the same order as in the list in section~\ref{sec_systematics}. 
For the combination procedures described in section~\ref{section_results},
the error sources marked by a $\dagger$ were considered to be $100\%$ correlated between the \Wln\ and \Wqq\ decay modes and also among the \com\ energies.
All other sources were assumed to be completely uncorrelated.  \label{table_systematic}}
\end{center}
\end{table}

\section{Results \label{section_results}}
\subsection{{\boldmath \W} polarisation}
Table~\ref{table_wpol} shows the fraction of longitudinally polarised \W\
bosons measured from the data samples at each nominal \com\ energy.
The values have been corrected for the detector effects described in
section~\ref{sec_detcorr} and the bias described in
section~\ref{sec_biascorr}.
The values obtained at $183$~\GeV\ and $189$~\GeV\ differ slightly from those
previously published by OPAL~\cite{opal_pr260,opal_pr323} due to the use of
improved MC generators (see section~\ref{sec_systematics}), to the inclusion of
the bias correction and to minor changes to the event reconstruction procedure.
For comparison with the data, table~\ref{table_wpol} also gives the fraction 
of longitudinally polarised \W\ bosons predicted by the \KANDY\ MC samples.
 
\begin{table}[h]
\begin{center}
\hspace*{-10pt}\begin{tabular}{|c||r|r|r|r|} \cline{2-5}
\multicolumn{1}{l|}{}&\multicolumn{4}{c|}{Longitudinal polarisation (\%)}\\
\hline
\multicolumn{1}{|c||}{$\sqrt{\mathit{s}}$}&\multicolumn{3}{c|}{OPAL
Data}&\multicolumn{1}{c|}{MC}\\ 
\multicolumn{1}{|c||}{(\GeV)}&\multicolumn{1}{c}{\Wln}&\multicolumn{1}{c}{\Wqq}& \multicolumn{1}{c|}{Combined}&\multicolumn{1}{c|}{Combined}\\ \hline
183&$18.8\pm10.3\pm1.1$ &$31.1\pm10.4\pm2.1$ &$24.8\pm7.4\pm1.4$&$26.4\pm0.2$\\ \hline
189&$17.6\pm5.8\pm1.0$ &$21.0\pm5.8\pm1.5$ &$19.2\pm4.3\pm1.1$&$25.6\pm0.2$\\ \hline
192&$56.7\pm14.3\pm0.9$&$21.7\pm13.8\pm1.7$&$38.6\pm9.7\pm1.1$&$25.2\pm0.2$\\ \hline
196&$16.8\pm8.3\pm0.9$ &$ 7.8\pm8.3\pm1.8$ &$12.4\pm5.9\pm1.1$&$23.9\pm0.2$\\ \hline
200&$32.5\pm8.4\pm1.1$ &$35.5\pm9.4\pm1.6$ &$33.9\pm6.2\pm1.1$&$23.2\pm0.2$\\ \hline
202&$30.4\pm11.8\pm1.6$&$33.7\pm12.5\pm1.8$&$31.9\pm8.9\pm1.3$&$23.2\pm0.2$\\ \hline
205&$33.1\pm8.7\pm1.4$ &$35.3\pm9.3\pm1.9$ &$34.1\pm6.4\pm1.4$&$22.2\pm0.2$\\ \hline
207&$17.5\pm6.4\pm1.2$ &$21.3\pm6.8\pm1.8$ &$19.2\pm4.8\pm1.3$&$21.7\pm0.2$\\ \hline\hline
Average&$23.6\pm2.9\pm0.7$&$24.5\pm3.0\pm1.7$&$23.9\pm2.1\pm1.1$&$23.9\pm0.1$\\ \hline
\end{tabular}
\caption{The fraction of longitudinal polarisation for the leptonically and
hadronically decaying \W\ bosons at each nominal \com\ energy after detector and bias corrections.
Also shown is the combined result obtained from the BLUE technique~\cite{blue} under the assumptions
described in section~\ref{section_results}.
The values extracted from the data are shown with both statistical and
systematic errors.
The values extracted from the generator-level \KANDY\ MC samples are shown with statistical errors only.
The last row of the table shows the luminosity-weighted averages.
 \label{table_wpol}}
\end{center}
\end{table}

The measured fractions obtained from the leptonic and hadronic decays were combined using the Best Linear Unbiased Estimator (BLUE) method~\cite{blue} in which each source of systematic error was assumed to be uncorrelated with all other
sources of systematic error and either $100\%$ or $0\%$ correlated between the
two decay modes (see table~\ref{table_systematic} for details). 
The correlations between the statistical errors for the two decay modes were
measured from the data at each \com\ energy and found to have a magnitude of
less than $10\%$.
The systematic and statistical errors were assumed to be uncorrelated with
each other.

Following the combination of the decay modes, the results from the eight \com\ energies were themselves combined to make a luminosity-weighted average in which the correlations between the systematic errors were again approximated as being either $100\%$ or $0\%$.
The average longitudinal polarisation was found to be
$(23.9\pm 2.1 \pm 1.1) \%$, which is in good agreement with the \KANDY~\cite{koralw,yfs} MC prediction of $(23.9\pm0.1)\%$. 

\begin{center}
\begin{figure}[p]
\begin{center}
\epsfig{file=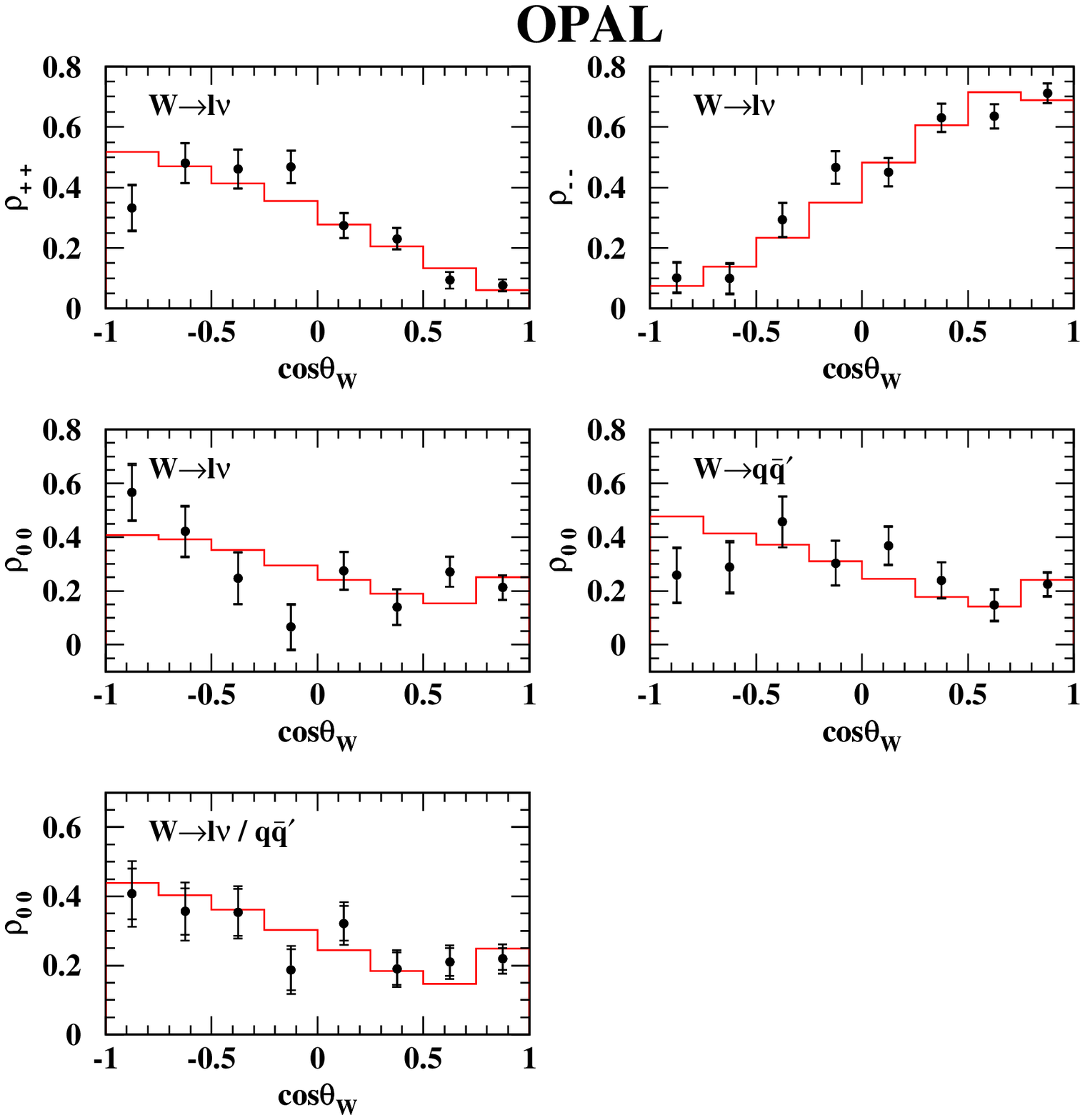,height=17cm}
\caption{Luminosity-weighted averages of the diagonal elements of the SDM as 
functions of \cthw.
The points show the data after detector and bias corrections.
The inner error bars show the statistical uncertainties and the outer error
bars show the total uncertainties including both the statistical and
systematic contributions.
The histograms show the generator-level \KANDY\ MC prediction.
The figures in the top row show \rpp\ and \rmm\ for leptonically decaying
\W\ bosons.
The figures in the middle row show \rzz\ for leptonically decaying \W\ bosons and hadronically decaying \W\ bosons separately.
The figure in the bottom row shows \rzz\ with the leptonic and hadronic decay
modes combined. 
By construction, the diagonal elements of the SDM are related to each other by 
the normalisation condition $\rpp+\rmm+\rzz=1$, but have not been individually constrained to the physically-allowed region between zero and one.  \label{figure_sdmsummary} }
\end{center}
\end{figure}
\end{center}

\begin{center}
\begin{figure}[p]
\begin{center}
\epsfig{file=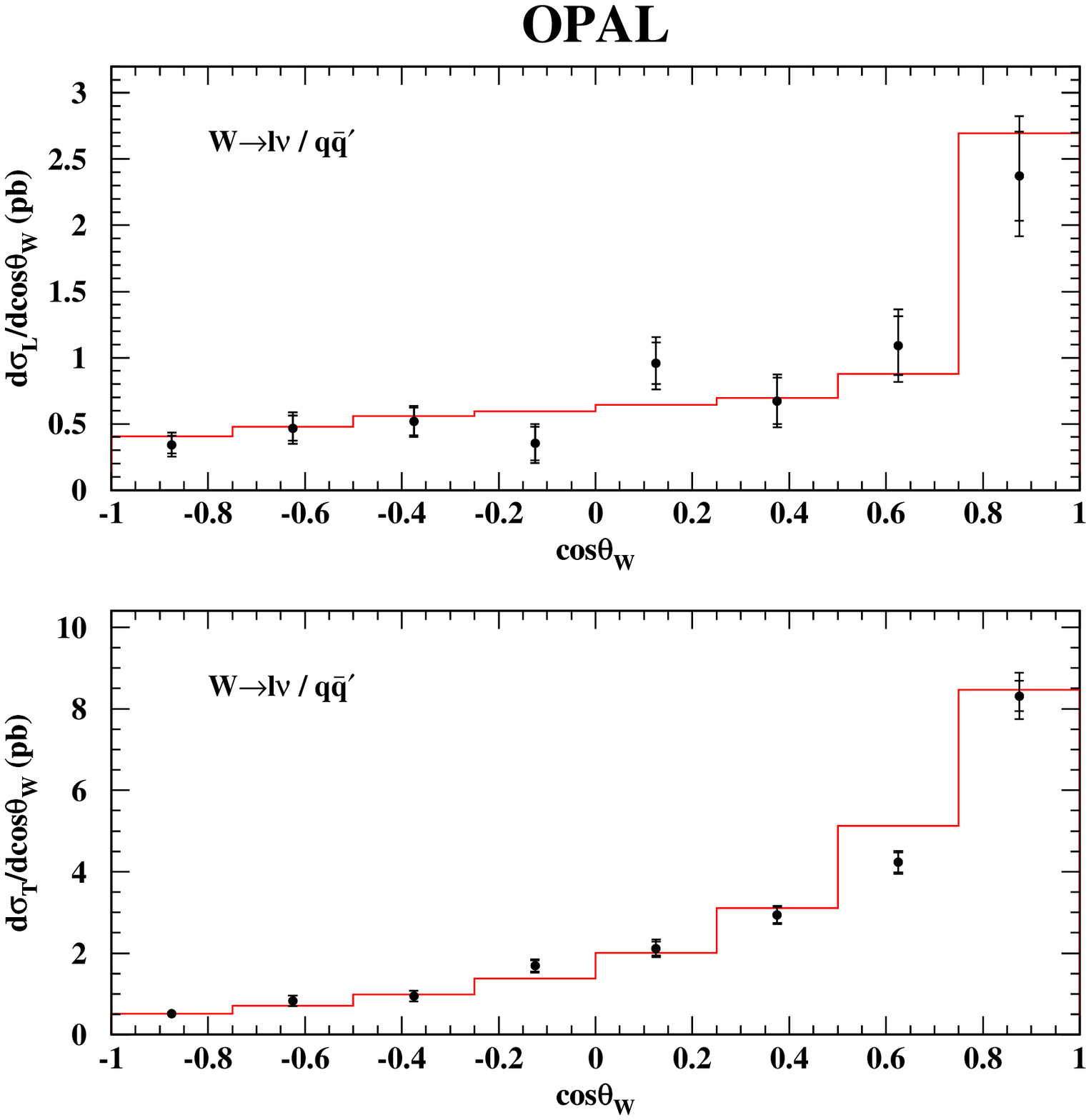,height=17cm}
\caption{The luminosity-weighted average polarised differential cross-sections
of section~\ref{subsection_polxs}, where the average is over the eight nominal
\com\ energies and over the \cthw\ bin width.
The points show the data after detector and bias corrections.
The inner error bars show the statistical uncertainties and the outer error
bars show the total uncertainties including both the statistical and
systematic contributions.
The histograms show the generator-level \KANDY\ MC prediction.
\label{figure_polsummary} }
\end{center}
\end{figure}
\end{center}

The luminosity-weighted averages of \rpp, \rmm\ and \rzz\ are shown in 
figure~\ref{figure_sdmsummary}.
The polarised differential cross-sections are shown in figure~\ref{figure_polsummary}.
The numerical values associated with the figures can be found
in~\cite{durham_database}.

\subsection{{\boldmath \CP}-violating effects}

\begin{center}
\begin{figure}[p]
\begin{center}
\epsfig{file=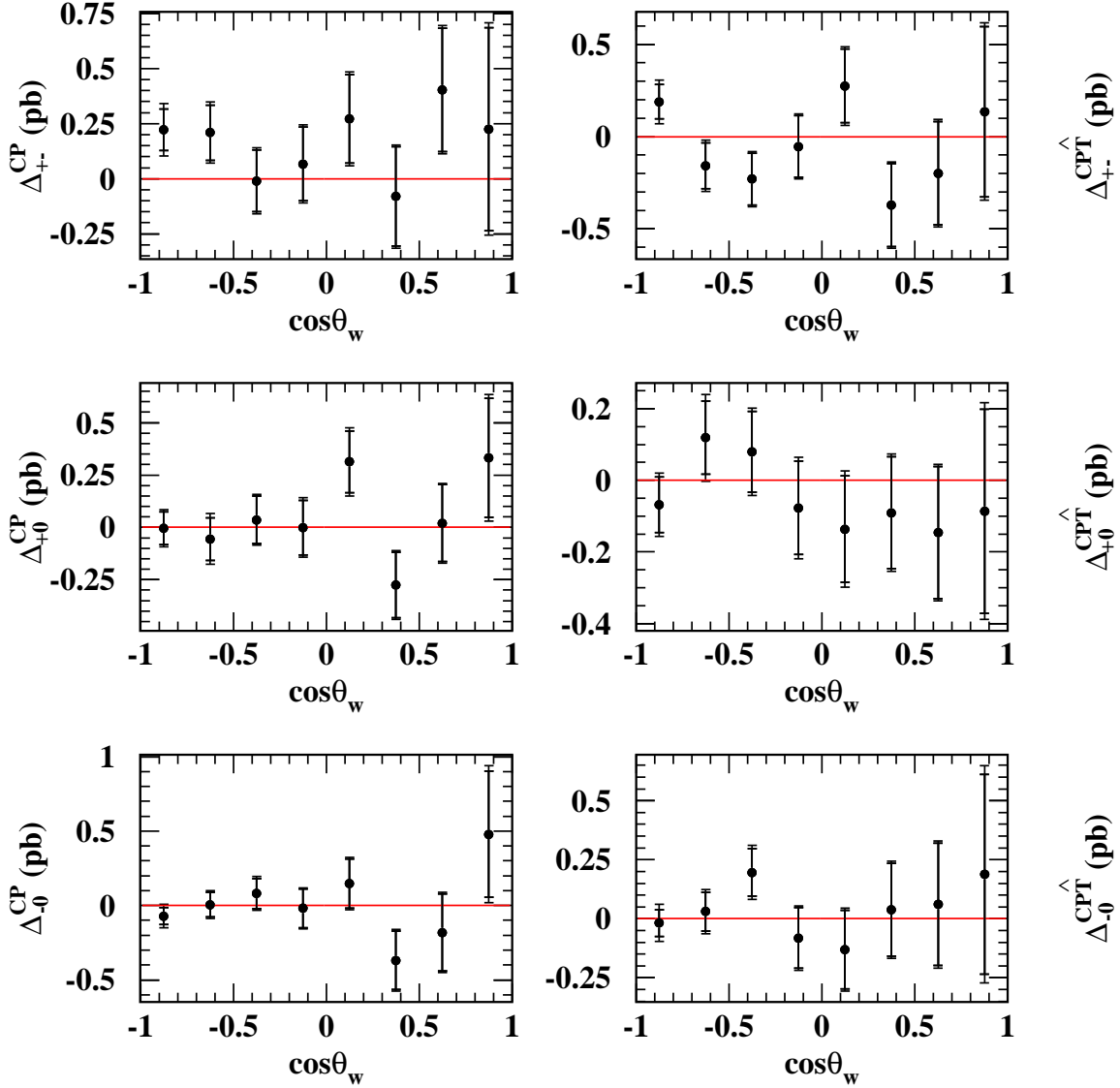,height=17cm }
\caption{ The luminosity-weighted average of the \CP-odd and \CPTh-odd observables of section~\ref{subsection_cptest}, where the average is over the eight nominal \com\ energies and over the \cthw\ bin width.
The observables are measured in units of picobarns and shown as functions of \cthw.
The points show the data after detector correction.
The inner error bars show the statistical uncertainties and the outer error
bars show the total uncertainties including both the statistical and
systematic contributions.
The solid lines show the \SM\ tree-level prediction.  \label{figure_cptests}}
\end{center}
\end{figure}
\end{center}

The luminosity-weighted averages of the \CP-odd and \CPTh-odd
observables derived from the \Wm\ and \Wp\ SDM elements in section~\ref{subsection_cptest} are shown in figure~\ref{figure_cptests} as functions of \cthw. 
The results are summarised in table~\ref{table_cpcptsum}.
Each source of systematic error in the off-diagonal elements of the SDM was
assumed to be 100\% correlated between the \Wm\ and \Wp\ results.
 
Any significant deviations from zero in the first row of
table~\ref{table_cpcptsum}
would constitute an unambiguous signature of \CP\ violation. 
Any significant deviations from zero in the second row of the table would
show the presence of loop effects.
Here, all results are consistent with the \SM\ tree-level prediction 
of zero within the quoted errors.

\begin{table}[h]
\begin{center}
\begin{tabular}{|c||r|r|r|} \cline{2-4}
\multicolumn{1}{c|}{}&
\multicolumn{1}{c|}{\rule[-1mm]{0mm}{6mm}$\Delta_{+-}$  (\pb)}&
\multicolumn{1}{c|}{\rule[-1mm]{0mm}{6mm}$\Delta_{+0}$  (\pb)}&
\multicolumn{1}{c|}{\rule[-1mm]{0mm}{6mm}$\Delta_{-0}$  (\pb)} \\ \hline
\rule[-1mm]{0mm}{6mm}\CP&$0.33\pm0.17\pm0.06$&$0.09\pm0.11\pm0.04$&$0.02\pm0.15\pm0.06$ \\ \hline
\rule[-1mm]{0mm}{6mm}\CPTh&$-0.10\pm0.17\pm0.06$&$-0.10\pm0.11\pm0.04$&$0.07\pm0.15\pm0.06$ \\
\hline
\end{tabular}
\caption{The luminosity-weighted average values of the \CP-odd and \CPTh-odd observables described in section~\ref{subsection_cptest} measured in picobarns.
The values extracted from the data are shown with both statistical and
systematic errors.
\label{table_cpcptsum}}

\end{center}
\end{table}

\section{Conclusion \label{section_conclusion}}

We have presented a spin density matrix analysis of \W\ bosons at LEP.
The diagonal elements of the SDM, the differential polarised cross-sections
and the fractions of longitudinal polarisation all show good agreement 
with the \SM\ prediction for each of the analysed data samples and for their
luminosity-weighted average.
Our result is also consistent with the fraction of
longitudinal polarisation recently measured by the L3 collaboration at LEP using
pairs of \W\ bosons decaying to the \qqen\ and \qqmn\ final states~\cite{L3}.

No evidence was found for \CP\ violation in \WW\ production although the
statistical precision of the analysis was not high enough to measure 
loop-level effects. 
The absence of \CP-violating effects in this model-independent study places
loose limits on the possible values of the \CP-violating \TGC\ parameters
($\tilde{\kappa}_{V}$, $\tilde{\lambda}_{V}$, $g^{4}_{Z}$)~\cite{YR_tgc},
which are constrained to be of the order $\Delta^{\CP}_{\tau\tau^{\prime}}$/$\sigma$~\cite{gounaris_cp}.
The \CP-odd observable $\Delta^{\CP}_{+-}$ showed the largest deviation from
zero with a luminosity-weighted average value of $0.33\pm0.17\pm0.06$
picobarns.
Using this result we find that the \CP-violating \TGC\ parameters are expected
to be less than or of the order $\mathcal{O}(10^{-1})$ in the \com\ energy range covered by this analysis. 

\medskip
\bigskip\bigskip\bigskip
\appendix
\par
\section*{Acknowledgements:}
\par
We particularly wish to thank the SL Division for the efficient operation
of the LEP accelerator at all energies
 and for their close cooperation with
our experimental group.  In addition to the support staff at our own
institutions we are pleased to acknowledge the  \\
Department of Energy, USA, \\
National Science Foundation, USA, \\
Particle Physics and Astronomy Research Council, UK, \\
Natural Sciences and Engineering Research Council, Canada, \\
Israel Science Foundation, administered by the Israel
Academy of Science and Humanities, \\
Benoziyo Center for High Energy Physics,\\
Japanese Ministry of Education, Culture, Sports, Science and
Technology (MEXT) and a grant under the MEXT International
Science Research Program,\\
Japanese Society for the Promotion of Science (JSPS),\\
German Israeli Bi-national Science Foundation (GIF), \\
Bundesministerium f\"ur Bildung und Forschung, Germany, \\
National Research Council of Canada, \\
Hungarian Foundation for Scientific Research, OTKA T-038240, 
and T-042864,\\
The NWO/NATO Fund for Scientific Research, the Netherlands.\\

 

\clearpage


\clearpage

\end{document}